\documentclass[aps,prd,preprint,superscriptaddress]{revtex4-1}
\usepackage[dvipdfmx]{graphicx}
\usepackage{CJKutf8}
\usepackage{xcolor}
\usepackage{url}
\usepackage{ulem}
\usepackage{lineno}

\newcommand{\Add}[1]{\textcolor{black}{#1}}
\newcommand{\Erase}[1]{}
\newcommand{\New}[1]{\textcolor{black}{#1}}

\begin{document}
\preprint{Paper-CANDLES-0nBB}

\title{Low background measurement in  CANDLES-III for studying the neutrino-less double beta decay of $^{48}$Ca}


\newcommand{\rcnp}{\affiliation{Research Center for Nuclear Physics (RCNP),
    Osaka University, Ibaraki, Osaka 567-0047, Japan}}
\newcommand{\osprc}{\affiliation{Project Research Center for Fundamental Science,
    Osaka University, Toyonaka, Osaka 567-0043, Japan}}
\newcommand{\osaka}{\affiliation{Graduate School of  Science, 
    Osaka University, Toyonaka, Osaka 560-0043, Japan}}
\newcommand{\tokushima}{\affiliation{ Department of Science and Technology, 
    Tokushima University, Tokushima, Tokushima 770-8506, Japan}}
\newcommand{\fukui}{\affiliation{Graduate School of Engineering, 
    University of Fukui, Fukui, 910-8507, Japan}}
\newcommand{\daisan}{\affiliation{Department of Environmental Science and Technology, 
    Osaka Sangyo University, Daito, Osaka 574-8530, Japan}}
\newcommand{\wakasa}{\affiliation{The Wakasa-wan Energy Research Center, 
    Tsuruga, Fukui 914-0192, Japan}}
\newcommand{\tsukuba}{\affiliation{
    Faculty of Pure and Applied Sciences, University of Tsukuba, Tsukuba, Ibaraki 305-8571, Japan}}


\author{S.~Ajimura}\rcnp
\author{W.~M.~Chan}\rcnp
\author{K.~Ichimura}\altaffiliation{Present address: Research Center for Neutrino Science, Tohoku University}\rcnp
\author{T.~Ishikawa}\rcnp
\author{K.~Kanagawa}\rcnp
\author{B.~T.~Khai}\rcnp
\author{T.~Kishimoto}\rcnp
\author{H.~Kino}\rcnp
\author{T.~Maeda}\altaffiliation{Present address: Sector of Fukushima Research and Development, Japan Atomic Energy Agency}\rcnp
\author{K.~Matsuoka}\rcnp
\author{N.~Nakatani}\rcnp
\author{M.~Nomachi}\rcnp
\author{M.~Saka}\rcnp
\author{K.~Seki}\rcnp
\author{Y.~Takemoto}\altaffiliation{Present address: Institute for Cosmic-ray Research, the University of Tokyo}\rcnp
\author{Y.~Takihira}\rcnp
\author{D.~Tanaka}\rcnp
\author{M.~Tanaka}\rcnp
\author{K.~Tetsuno}\rcnp
\author{V.~T.~T.~Trang}\rcnp
\author{M.~Tsuzuki}\rcnp
\author{S.~Umehara}\rcnp

\author{K.~Akutagawa}\osaka
\author{T.~Batpurev}\osaka
\author{M.~Doihara}\osaka
\author{S.~Katagiri}\osaka
\author{E.~Kinoshita}\osaka
\author{Y.~Hirano}\altaffiliation{Present address: Graduate School of Medicine, Nagoya University}\osaka
\author{T.~Iga}\osaka
\author{M.~Ishikawa}\osaka
\author{G.~Ito}\osaka
\author{H.~Kakubata}\osaka
\author{K.~K.~Lee}\osaka
\author{X.~Li}\osaka
\author{K.~Mizukoshi}\altaffiliation{Present address: Department of Physics, Kobe University}\osaka
\author{M.~Moser}\osaka
\author{T.~Ohata}\osaka
\author{M.~Shokati}\osaka
\author{M.~S.~Soberi}\osaka
\author{T.~Uehara}\osaka
\author{W.~Wang}\osaka
\author{K.~Yamamoto}\osaka
\author{K.~Yasuda}\osaka
\author{S.~Yoshida}\osaka
\author{N.~Yotsunaga}\osaka

\author{T.~Harada}\fukui
\author{H.~Hiraoka}\fukui 
\author{T.~Hiyama}\fukui 
\author{A.~Hirota}\fukui 
\author{Y.~Ikeyama}\fukui 
\author{A.~Kawamura}\fukui 
\author{Y.~Kawashima}\fukui 
\author{S.~Maeda}\fukui
\author{K.~Matsuoka}\fukui
\author{K.~Nakajima}\fukui
\author{I.~Ogawa}\fukui
\author{K.~Ozawa}\fukui
\author{K.~Shamoto}\fukui
\author{K.~Shimizu}\fukui
\author{Y.~Shinki}\fukui
\author{Y.~Tamagawa}\fukui
\author{M.~Tozawa}\fukui 
\author{M.~Yoshizawa}\fukui

\author{K.~Fushimi}\tokushima
\author{R.~Hazama}\daisan
\author{P.~Noithong}\daisan
\author{A.~Rittirong}\daisan
\author{K.~Suzuki}\wakasa
\author{T.~Iida}\tsukuba


\collaboration{CANDLES Collaboration}\noaffiliation

\date{\today}

\begin{abstract}
\Add{We developed a CANDLES-III system to study the neutrino-less double beta (0$\nu\beta\beta$) decay of $^{48}$Ca.
The proposed system employs 96 CaF$_{2}$ scintillation crystals~(305\,kg) with natural Ca~($^{\rm nat.}$Ca) isotope which corresponds 350\,g of $^{48}$Ca.}
\Add{External backgrounds were rejected using a 4$\pi$ active shield of a liquid scintillator surrounding the CaF$_2$ crystals.}
The internal backgrounds caused by the radioactive impurities within the CaF$_2$ crystals can be reduced effectively through analysis of the signal pulse shape.
We analyzed the data \Erase{observed}\Add{obtained} in the Kamioka underground for a live-time of 130.4 days to evaluate the feasibility of the low background measurement with the CANDLES-III detector.
\Erase{Moreover, we}\Add{Using Monte Carlo simulations, we} estimated \Erase{the number of background events from the simulation based on the radioactive impurities in the CaF$_{2}$ crystals and the rate of high energy $\gamma$-rays caused by the (n, $\gamma$) reactions induced by environmental neutrons.}\Add{the background rate from the radioactive impurities in the CaF$_{2}$ crystals and the rate of high energy $\gamma$-rays caused by the (n, $\gamma$) reactions induced by environmental neutrons.}
The expected background rate was in a good agreement with the measured rate\Add{, i.e., approximately 10$^{-3}$~events/keV/yr/(kg of $^{\rm nat.}$Ca), in the 0$\nu\beta\beta$ window.}
In conclusion, the background candidates were estimated properly by comparing the measured energy spectrum with the background simulations.
With this measurement method, we performed the first search for 0$\nu\beta\beta$ decay in a low background condition using a detector on the scale of hundreds of kg of non-enriched Ca. 
Deploying scintillators enriched in $^{48}$Ca will increase the sensitivity strongly. 
$^{48}$Ca has a high potential for use in 0$\nu\beta\beta$ decay search, and is expected to be useful for the development of a next-generation detector for highly sensitive measurements.
\end{abstract}


\maketitle


\section{Introduction}
\Add{The origin of neutrino masses and the absolute mass scale of neutrinos are major open questions related to neutrino properties.}
These questions can potentially be investigated by observing the neutrino-less double beta decay (0$\nu\beta\beta$ decay)~\cite{Furry1939}.
Double beta ($\beta\beta$) decay has two modes.
One of these modes is the two-neutrino $\beta\beta$ decay (2$\nu\beta\beta$ decay) accompanied by two electrons and two anti-neutrinos~\cite{Mayer1935}.
This decay mode is allowed within the standard model of particle physics
and has been observed in several isotopes~\cite{Barabash2010}\cite{Barabash2019}. 
\Add{The other is 0$\nu\beta\beta$ decay, which can occur only if neutrinos are Majorana particles~\cite{Majorana1937}.} 
This process is forbidden in the standard model because it violates the lepton number conservation law.
The half-life of this decay is inversely proportional to the square of the effective Majorana neutrino mass,
\Add{under the assumption that 0$\nu\beta\beta$ decay occurs via the exchange of a light Majorana neutrino}~\cite{Schechter}\cite{Ejiri2019}\cite{Ejiri2019f}.

The 0$\nu\beta\beta$ decay has been searched for several isotopes~\cite{Dolinski}, but has not yet been observed.
\Add{Recent experiments reported lower limits on the half-lives of
T$^{0\nu}_{1/2}$ $\ge$ 1.07$\times$10$^{26}$\,yr (KamLAND-Zen~\cite{KamLAND-Zen2016}, $^{136}$Xe),
3.5$\times$10$^{25}$ yr~(EXO-200~\cite{EXO2019}, $^{136}$Xe),
1.8$\times$10$^{26}$ yr~(GERDA~\cite{GERDA2020}, $^{76}$Ge),
1.9$\times$10$^{25}$ yr~(MAJORANA~\cite{Majorana2018}, $^{76}$Ge),
and 3.2$\times$10$^{25}$ yr~(CUORE~\cite{CUORE2020}, $^{130}$Te) at a 90\% confidence level.}

The T$^{0\nu}_{1/2}$ sensitivity scales linearly with the source mass ($M$) and measurement time ($t$) in a background-free case, as opposed to $\sqrt{Mt}$ \Add{in a background-dominated case.}
Thus, a background-free experiment is necessary to \Add{efficiently} \Add{improve the sensitivity through the }scaling up of the source \Erase{volume}\Add{mass} and measurement time.

\Erase{In this respect, the search for the 0$\nu\beta\beta$ decay using the}\Add{The} $^{48}$Ca isotope is particularly promising because it has the highest Q-value\\ (Q$_{\beta\beta}$=4267.98$\pm$0.32~keV~\cite{48Ca-Qvalue}) among known $\beta\beta$ decaying isotopes~\cite{DBD-Isotope-Table_Zdesenko}.
This Q$_{\beta\beta}$-value is above the bulk of natural radioactive backgrounds \Add{where the maximum energies of $\gamma$-rays and $\beta$-rays are 2.61 MeV from $^{208}$Tl decay and 3.27 MeV from $^{214}$Bi decay, respectively.} \Erase{and ensures a favorable phase space that enhances the 0$\nu\beta\beta$ decay rate.}
Thus, background candidates are expected to be limited, and low background measurement can be realized.
\Add{The large Q$_{\beta\beta}$-value also ensures a favorable phase space that enhances the 0$\nu\beta\beta$ decay rate.}

The double magic number nucleus $^{48}$Ca is the lightest such nucleus used in 0$\nu\beta\beta$ search experiments thus far and is considered to be an interesting target for nuclear theory. 
This nucleus has a relatively clear and simple shell structure with a reasonable number of nucleons. 
It is suitable for the calculation of nuclear matrix element~(NME) by shell model~\cite{PhysRevLett.116.112502} and also an ideal 'reference' for comparing the different models.
All the models for NME calculations may also be applied to the heavier $\beta\beta$ nuclei, but at an expense of validity due to other restrictions, e.g. in configuration space or number of nucleons.

\Add{The natural abundance of $^{48}$Ca is known to be \Erase{the least}\Add{relatively small}, at approximately 0.2\%, among $\beta\beta$ decaying nuclei.
Several $^{48}$Ca enrichment techniques are currently under development to overcome this disadvantage~\cite{Elemag2001,LIS2020,Umehara-PTEP2015,Hazama-KURRI,Kishimoto-PTEP2015}.}

Searches for the 0$\nu\beta\beta$ decay of $^{48}$Ca were first demonstrated approximately 60 years ago~\cite{McCarthy1955}.
Since then, although a variety of measurements have already been performed~\cite{48Ca1966,48Ca1970,48Ca-Beijing, 48Ca-TGV, Ogawa2014, Umehara2008, NEMO3-48Ca-2016}, no signals for 0$\nu\beta\beta$ decay were observed in any of these attempts.
The best limit on $^{48}$Ca is currently set by the ELEGANT VI experiment, which used 6.6 kg of \Add{$^{\rm nat.}$CaF$_{2}$(Eu) scintillators}
(7.6 g of $^{48}$Ca) at T$^{0\nu}_{1/2}$ $\ge$ 5.8$\times$10$^{22}$ yr~\cite{Umehara2008}.
The measurement was not limited by the backgrounds.
The \Add{differentiating}\Erase{characteristic} of the ELEGANT VI detector was the usage of a 4$\pi$ active shield by its scintillator complex, the success of which was the key to achieving a background-free measurement.

Our strategy for achieving better sensitivity is to increase \Erase{a number of target nuclei}\Add{the source mass} and to maintain the background at a lower level \Add{as that of the ELEGANT VI experiment}, for which the initial concept of the CANDLES system is proposed.
The Eu-doped CaF$_2$ used in ELEGANT VI has a large light output and provides a good energy resolution. 
However, it has a short attenuation length because of the self-absorption of its scintillation light and is not suitable for the next step.
A scale up with a good energy resolution without degradation of the light collection can be achieved via utilization of a combination of undoped CaF$_2$ with a long attenuation length (i.e., more than 10~m) and a layer of the adjacent wavelength shifter (WLS)~\cite{Yoshida2009}.
We realize a 4$\pi$ active shield by surrounding the CaF$_2$ crystals in all directions with a luminous liquid scintillator (LS) to accomplish background-free measurement.
The high Q$_{\beta\beta}$-value of $^{48}$Ca combined with the 4$\pi$ active shield achieved a substantial reduction in the backgrounds.

This article presents a method for achieving a low background condition. 
\Add{In Chapter II, we describe our detector.
Chapter III details the analytical parameters that have an important role in background analysis.
In Chapter IV, we report the methods and our first result of CANDLES-III in the Kamioka underground, which is comparable to the best result among 0$\nu\beta\beta$ measurements with $^{48}$Ca.}
\Add{In Chapter V, we discuss results related to the observed background rate and the future prospects of our method, and in Chapter VI, we summarize our study.}

\section{CANDLES-III detector}
\label{sec:detector}
\begin{figure}[htbp]
\begin{center}
\includegraphics[width=1.0\linewidth,pagebox=cropbox,clip]{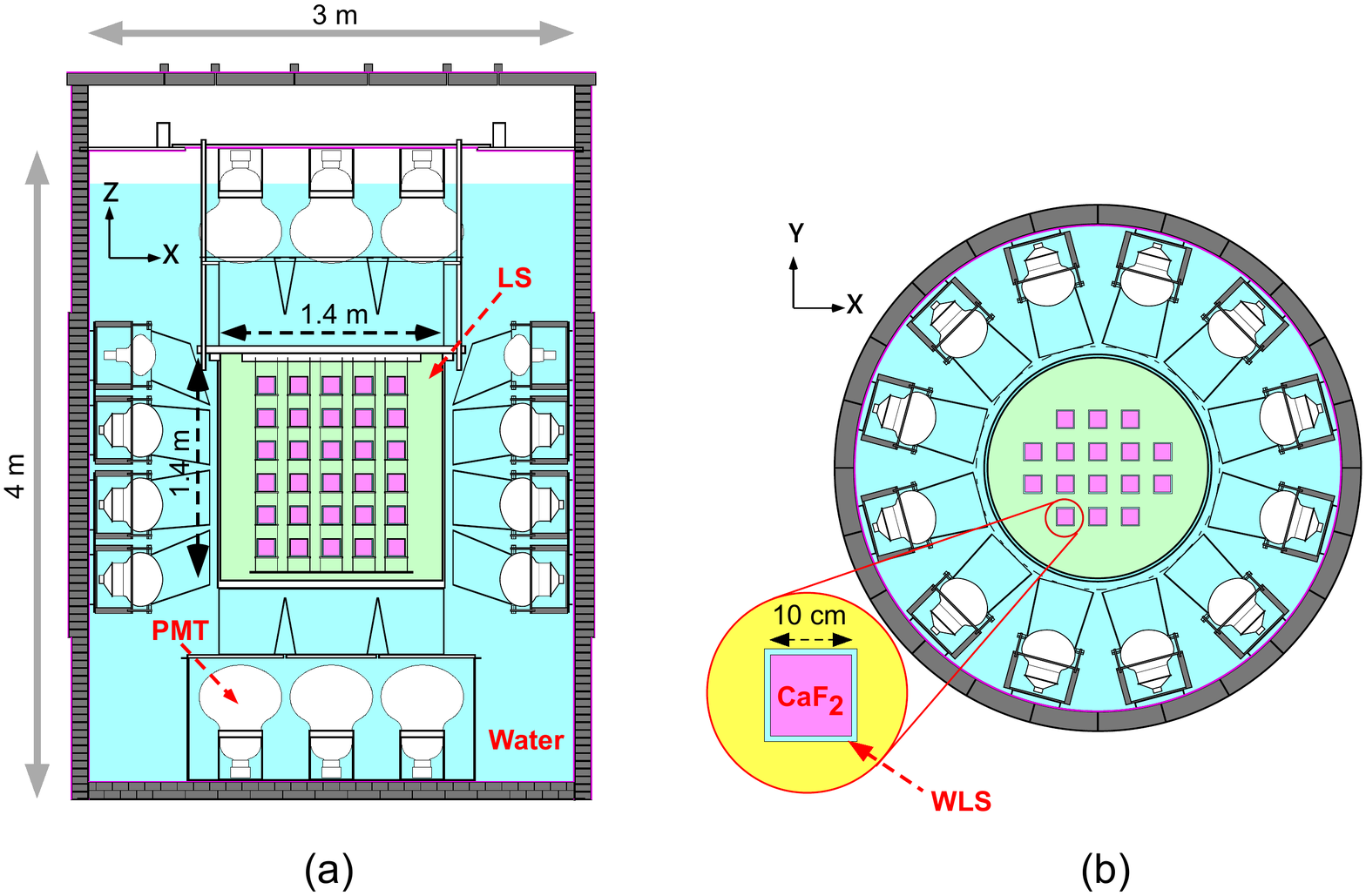}
\caption{Detector setup of CANDLES-III (a: side view and b: top view).
\Add{Magnified}\Erase{A zoomed} figure in (b) shows details of CaF$_2$ module.
The layer number ($N_l=$ 1 -- 6) is from top to bottom.
The module number ($N_m=$ 1 -- 96) first increases in X-axis, and then in negative Y and Z-axes.
For top layer ($N_l=1$), $N_m=$ 1 -- 16. For bottom layer ($N_l=6$), $N_m=$ 81 -- 96.
Here, $N_m$ of zoomed CaF$_2$ module is 14.}
\label{Fig:Detector}
\end{center}
\end{figure}

CANDLES (CAlcium fluoride for the study of Neutrinos and Dark matters by Low Energy Spectrometer) is an experiment performed to search for the 0$\nu\beta\beta$ decay of $^{48}$Ca using undoped \Add{$^{\rm nat.}$CaF$_2$} (CaF$_2$) scintillation crystals.
The conceptual design of the CANDLES detector is described in Ref.~\cite{Yoshida2009}.
The current setup of the detector system, which is called CANDLES-III, is installed at lab-D of the Kamioka Underground Laboratory (2700 m.w.e.) in the Institute of Cosmic Ray Research of the University of Tokyo.
Fig.~\ref{Fig:Detector} shows the CANDLES-III system, which comprises 96 CaF$_2$ modules, a LS, a water buffer, 62 photomultiplier tubes (PMTs), and external shields.

The CaF$_2$ module consisted of a \Erase{10~cm}\Add{(10\,cm)$^3$} cube (3.18~kg) undoped CaF$_2$ crystal, a 5 mm-thick layer of WLS phase, and a 3 mm-thick acrylic container.
In the WLS phase, the emission light of CaF$_2$ with its peak in the ultraviolet (UV) region was immediately converted to visible light, where the quantum efficiency of the PMTs was sufficient (maximum; $\sim400$~nm), and the materials in the optical path had good transparencies~\cite{Yoshida2009}.
The structure of the WLS phase surrounding the CaF$_2$ crystals was essential because the LS was not transparent to UV light.
The WLS phase was composed of mineral oil in which WLS (Bis-MSB; 0.1 g/$\ell$)~\Add{\cite{Yoshida2009}} was dissolved.
A total of 96 CaF$_2$ modules were arranged in six horizontal layers in the vertical direction, such that 16 CaF$_2$ modules were positioned on each layer in an acrylic, cylindrical tank (LS vessel)\Erase{ of a cylindrical shape of} 1.4~m in diameter and 1.4~m in height.
The layer number $(N_l)$ and module number $(N_m)$ are defined in the Fig.~\ref{Fig:Detector} caption.
The CaF$_2$ modules were suspended by wires from the ceiling of the LS vessel.
Relatively radio-pure crystals\Add{, in which $^{232}$Th series impurities were less than 10~$\mu$Bq/kg,} were assembled in the central part of the detector based on the results of the commissioning runs~\cite{Umehara:2015lla,Nakajima:2015yla,Iida:2016vfi,Iida:2016yct,Umehara2016}.
The total mass of the 96 CaF$_2$ crystals was 305~kg, which included 350~g of $^{48}$Ca.

The LS vessel was filled with the LS, which consisted of 80\% mineral oil, 20\% pseudo-cumene, and WLSs (PPO; 1.0~g/$\ell$ and Bis-MSB; 0.1 g/$\ell$)~\Add{\cite{Yoshida2009}}.
The LS was used as an active shield, as will be described later.
Prior to the commissioning run, we performed LS purification via both liquid--liquid extraction and nitrogen purge.

The scintillation lights from the CaF$_2$ modules and the LS were viewed using 62 PMTs (20-inch $\times$ 14, 13-inch $\times$ 36, and 10-inch $\times$ 12).
\Add{In this study, the definition of energy was determined based on the light yield of the CaF$_2$ crystals.
Even if \Add{the energy loss in LS was the same as the one in CaF$_2$}\Erase{there was the same energy loss as CaF$_2$ in LS}, this value was added as the energy equivalent to the light yield of CaF$_2$. 
The relative light yield of LS was approximately half that of CaF$_2$, which was dependent on the energy deposit.}

Fig.\ref{Fig:Detector} illustrates the configuration for the three types of PMTs.
\Add{Twenty-inch~(R7250MOD) PMTs were installed on the top and bottom, and 13-inch~(R8055H) and 10-inch~(R7081MOD) PMTs were installed on the side in 4 rows and 12 columns.
The 10-inch PMTs were used on the top row and the 13-inch PMTs were used on the other rows.
All PMTs and related accessories were provided by Hamamatsu Photonics K.K..}
A light collection system was installed between the PMTs and the LS vessel to improve the scintillation light collection~\cite{Umehara:2015lla}.
The LS vessel and PMTs were installed in a stainless-steel water tank of 3~m in diameter and 4~m in height.
We employed cancellation coils outside the external shield to reduce the Earth's magnetic field, \Add{which would otherwise affect the charge collection efficiency of the large-diameter PMTs~\cite{SUZUKI1993299}.}
The coils were covered with a flame-retardant material. 
We also adjusted the coil current to reduce the Earth's magnetic field of $\sim400$~mG to below 60~mG in average at every PMT position, based on a magnetic field simulation.
A high-voltage power supply for the PMTs and current supplies for cancellation coils with an interlock were installed in the data acquisition~(DAQ) hut.

\Erase{The whole detector was covered with external passive shields to reduce the high energy $\gamma$-rays produced by the environmental neutron capture reactions~((n,$\gamma$) reaction) in the detector materials and rocks~\cite{CANDLES-ngammaBG}.}

We employed a Pb shield (gray bricks, Fig.~\ref{Fig:Detector}) outside of the water tank for the $\gamma$-rays produced by environmental neutron capture reactions~((n,$\gamma$) reaction) on the rocks~\cite{CANDLES-ngammaBG}.
The typical Pb thickness was denoted to be 10~cm, to reduce $\gamma$-rays with several MeVs by 1/100.
For a more effective reduction of the external background, the Pb thickness was constructed to be 12~cm in the center of the detector side, where the passive water shield was relatively thin.

The stainless-steel water tank was another source of high energy $\gamma$-rays.
Accordingly, Si rubber sheets containing 40 wt.\% of \Add{$^{\rm nat.}$B$_4$C} (B sheet, purple sheets shown in Fig.~\ref{Fig:Detector}) were attached to both the inside and outside of the tank to prevent (n,$\gamma$) reaction inside the Pb shield.
The B sheet thickness of 4~mm was enough to reduce the thermal neutrons by approximately 1/1000.

A detector cooling system was installed to increase the scintillation light output emitted by the CaF$_2$. 
The light output from the \Erase{undoped} CaF$_2$ increased by 2\% with a temperature decrease of 1 $^\circ$C at room temperature~\cite{CaF2-tempdependence}.
The lab-D temperature was cooled down to \Add{a few $^\circ$C by an air conditioner}\Erase{a temperature slightly higher than the freezing point of water used as the passive shield (2~$^\circ$C for the setting temperature)}.
\Add{In addition, the water in the tank was circulated through a chiller to cool the detector to a temperature slightly above the freezing point of the water (2~$^\circ$C for setting temperature).}
\Add{The water temperature inside the tank measured by the thermistor was stable at 4.50$\pm$0.05~$^\circ$C, and once thermal equilibrium was reached, 
the whole detector was considered to be equally stable.}
\Erase{Consequently, the center of the detector was cooled down to 4.5~$^\circ$C and stabilized within $\pm$0.05~$^\circ$C.}
The light output of the CaF$_2$ was increased by approximately 30\% compared to that at room temperature.
\Erase{The stabilized temperature led to no detectable change of gain caused by the temperature~\cite{Iida:2016yct}.}
\Add{The stabilized temperature led to no detectable change in gain~\cite{Iida:2016yct}.}

For long-term stable measurement, a system for monitoring the laboratory environment (e.g., temperature, humidity, atmospheric pressure in lab-D, water temperatures inside the water tank, temperature and humidity in the DAQ hut) was installed, and the data were continuously recorded.
The liquid levels of water and LS were constantly monitored. 
We also installed leak detectors for water and LS and connected their alarm signals to the interlock of the power supplies for the safe operation of the detector, which contains a flammable substance (i.e., LS).

The DAQ system consisted of hardware that included Flash ADCs (FADCs), trigger logic on the $\mu$TCA \Add{(micro Telecom Computing Architecture)} system, and network DAQ software.
The signal waveform for each PMT was recorded for approximately 9 $\mu$sec by an 8-bit 500~MHz FADC~\cite{uTCA-khai}.
The waveform was recorded in a 2~nsec time-bin for the first 768~nsec and in a 64~nsec time-bin, which was the sum of 32 samples of 2~nsec time-bin data, for 8.2~$\mu$sec with time stamps.
It was read through a SpaceWire datalink~\cite{SpW} on the $\mu$TCA backplane.
The pulse shape digitized from the analog waveform was used in an offline analysis of the background reduction through pulse shape discrimination.

We developed a dual-gate trigger system to efficiently collect the long decay-time signals by rejecting the LS signals with short decay-times~\cite{TRG-maeda}.
The signals were integrated into two different time windows (“dual gate”) in the FPGA, and triggered only when both integrated values exceeded each threshold.

Random trigger events via a pulser were acquired at 3 Hz.
Single photoelectron events that accidentally entered these random trigger events were then collected. 
The charge-photoelectron conversion coefficient was calculated for each PMT every 24 hours based on the average value of the FADC counts of these single photoelectron events.

The energy threshold was set to tag the $\alpha$ decay of $^{212}$Bi $\rightarrow$ $^{208}$Tl.
This 6.05~MeV $\alpha$-ray caused a scintillation that was of the same amount as that of the 1.63~\Erase{MeV}\Add{MeV$_{\rm ee}$ (in electron equivalent energy)} $\beta$-ray because of CaF$_2$ scintillation quenching.
The energy thresholds for each CaF$_2$ module were distributed between 0.8 and 1.2~MeV (Fig.~15 in Ref.~\cite{TRG-maeda}) to detect the $\alpha$-ray of $^{212}$Bi and to identify subsequent $^{208}$Tl decay.

The software we developed herein was based on a system that uses a DAQ-Middleware framework and an online monitoring system~\cite{DAQ-suzuki}\cite{DAQ-MW}.
The FADC data were collected by fast reader components installed in a personal computer (PC) via SpaceWire to a Gigabit Ethernet interface.
The slow data (H.V. values, temperatures, etc.) were collected by slow reader components installed in another PC.
The data were partly analyzed for the online monitoring system and fully recorded to a hard disk for offline analysis.

\section{Parameters and detector performance for data analysis}
\label{sec:experiment}
\Add{The characteristic of the 0$\nu\beta\beta$ signal in CANDLES-III is the signal of the $\beta$-rays alone (i.e., without LS emission) in a single CaF$_2$ crystal, and has an energy equivalent to the Q-value. 
In section A, we will describe pulse shape discrimination~(PSD) analysis for identifying the signals of $\beta$-rays from those of $\alpha$-rays and LS; in section B, we will explain position reconstruction to determine the crystal in which the event deposits energy; and in section C, we will discuss energy reconstruction and calibration.}
\subsection{Pulse shape analysis}
\label{sec:psd}
\Erase{In the analysis of the CANDLES-III experiment, pulse shape discrimination~(PSD) parameters were used to discriminate the following $\beta$-rays and $\gamma$-rays events that deposited energy only in the CaF$_2$ crystals ($\beta$-events); events involving  LS emission ($\beta$+LS-events) to realize a 4$\pi$ active shield; and $\alpha$-ray events in CaF$_2$ ($\alpha$-events) for the background rejection analysis.}
In the analysis of the CANDLES-III experiment, we used \Erase{pulse shape discrimination~(PSD)}\Add{PSD} parameters to discriminate the $\beta$-rays and $\gamma$-rays events that deposited energy only in the CaF$_2$ crystals ($\beta$-events); events involving  LS emission ($\beta$+LS-events); and $\alpha$-ray events in CaF$_2$ ($\alpha$-events).

\begin{figure}[h!]
\begin{center}
\includegraphics[width=1.00\linewidth]{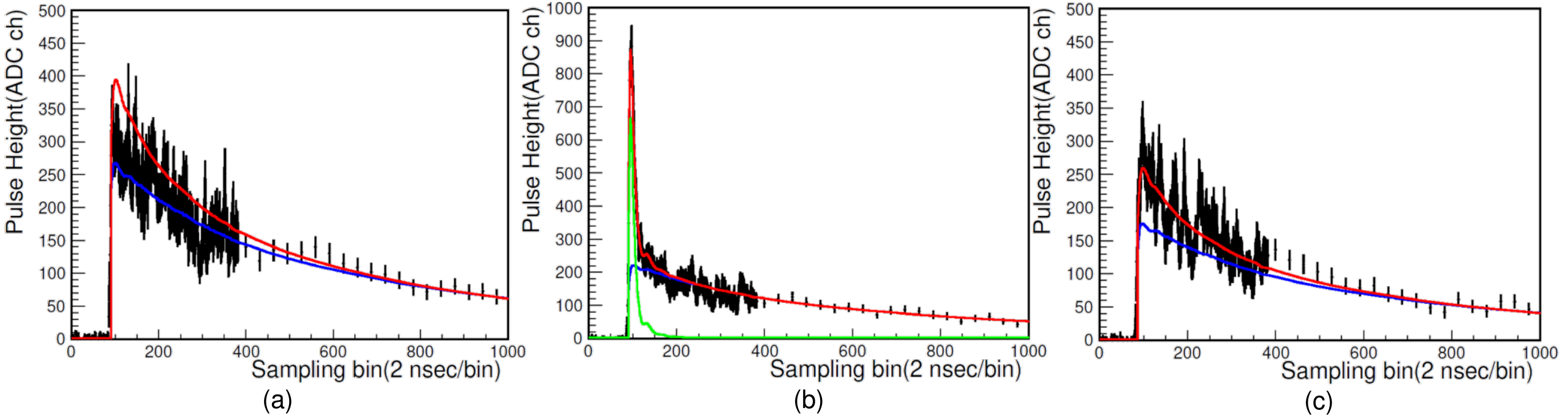}
\caption{Scintillating pulse shape observed using CANDLES-III detector. 
Recorded waveforms for 62 PMTs are summed up.
Data were recorded in 2~nsec time-bin for earlier 768~nsec region and 64~nsec time-bin for later region. 
\Erase{The energy deposits of both events are $\sim$2.6~MeV.}
\Add{(a) Observed pulse shapes for $\beta$-event, 
(b) $\beta$+LS-event, and (c) $\alpha$-event.
Blue, red, and green lines show reference pulse shapes of $\beta$-, $\alpha$-, and LS-events, respectively.}
\label{Fig:PulseShape}}
\end{center}
\end{figure}

The LS acted as \Erase{an} \Add{a 4$\pi$} active shield for the immersed CaF$_{2}$ modules to reduce the background events accompanied by $\gamma$-rays.
\Erase{The active shield was achieved by observing the pulse shapes.
The decay-time of CaF$_{2}$ scintillation was 1~$\mu$sec, whereas that of the LS was typically a few tens of nanoseconds.}
The events from the 0$\nu\beta\beta$ decay led to energy deposits only in the CaF$_{2}$ and were identified \Add{as $\beta$-events}, whereas the background events accompanied by $\gamma$-rays were able to yield energy deposits in both the LS and CaF$_{2}$ (and identified as $\beta$+LS-events).
The pulse shape of the $\beta$+LS-events had a large prompt component, whereas that of the \Erase{0$\nu\beta\beta$ candidate signals}\Add{$\beta$-events} did not, as shown in Figs.~\ref{Fig:PulseShape}(a) and \ref{Fig:PulseShape}(b).
\Erase{Thus, by observing the pulse shapes, the background signals of the $\beta$+LS-events can be discriminated as yielding a certain amount of energy deposits in the LS.}

The other major backgrounds in CANDLES-III were ${}^{212}$Bi$\rightarrow{}^{212}$Po$\rightarrow{}^{208}$Pb sequential decay and $^{208}$Tl decay within the CaF$_{2}$ crystal, as described in detail in Section~\ref{sec:bg-candidate}.
\Erase{Alpha-rays were involved in these background reduction because $^{212}$Po emitted an $\alpha$-ray, and the $^{208}$Tl decays were proceeded through the $\alpha$-ray by the $^{212}$Bi decay, as described detail in Section~\ref{sec:bg-candidate}.
Thus, identifying $\alpha$-events is important for the background reduction.}
\Erase{Alpha-rays were involved in these backgrounds, as described in detail in Section~\ref{sec:bg-candidate}. }
The $^{212}$Po decay emitted an $\alpha$-ray. The $^{208}$Tl decay followed the $^{212}$Bi $\alpha$ decay. 
Therefore, it was important for the background rejection to identify the $\alpha$-events.
\Erase{CaF$_2$}\Add{We} can discriminate between $\beta$- and $\alpha$-events based on the characteristics of the prompt part of the signal pulse shape~\cite{Umehara:2015lla}\Erase{.}\Add{, as shown in Figs.~\ref{Fig:PulseShape}(a) and \ref{Fig:PulseShape}(c).}

\subsubsection{Reference pulse shape for PSD}
The PSD analysis was performed via comparison of the observed pulse shape of each event with reference pulse shapes.
The three types of average pulse shape \Erase{of the $\beta$-events in the CaF$_2$ crystal, LS-events, and $\alpha$-events}created as reference, namely, the reference pulse shapes for the $\beta$-, LS-, and $\alpha$-events, are depicted in Fig.\ref{Fig:PulseShape} using blue, green, and red lines, respectively.

The reference pulse shape of the $\beta$-events was made using 2.615 MeV $\gamma$-rays from $^{208}$Tl decays that deposited energy only in the CaF$_2$ crystal.
The events yielding 2.615~MeV energy deposits were caused by $\gamma$-rays from $^{208}$Tl decays outside the LS vessel (PMT, etc.).\Erase{In other words,} \Erase{The $\gamma$-rays always passed through the LS and frequently deposited energy in the LS.}\Erase{The light yield of the LS was smaller than that of the CaF$_2$ module; thus}
\Add{The energy was calibrated for the light yield of CaF$_2$. 
Since the corresponding light yield of LS was approximately half that of the CaF$_2$,}
the multiple scattering events in both the CaF$_2$ and LS were distributed \Add{in} a lower energy region compared to the peak of the $\gamma$-rays (2.615 MeV).
The events in the energy window between 2.64 and 2.77 MeV were used to select $\beta$-events that had negligible energy deposits in the LS.
We evaluated biases in creating the reference pulse shape for the $\beta$-events, where only the events in the higher energy region of the $\gamma$-ray peak were selected.
The pulse shape from CaF$_2$ alone was obtained using the detector setup without LS installation. 
The reference pulse shape was created by $\gamma$-rays from $^{208}$Tl decays selected based on the same cut condition in the full CANDLES-III setup.
The obtained average pulse shape without the LS was consistent with the reference pulse shape created using the method described previously.
The light yield, light collection, and detector temperature of the setup without the LS were different from those of the full setup.
Therefore, the reference pulse shape of $\beta$-events in the full setup was used for the PSD analysis.

\Add{Events} with energy deposits above 300~keV only in the LS were used to create the reference pulse shape for the  LS-events.
\Add{In addition}, only the events near the detector center\Add{ ($\pm$30~cm from the detector center)}
 were selected.
\Add{because the recorded pulse height of the PMT near the position of the LS-event overflowed because of the short decay-time of the LS signal.}
The LS-events were obtained by different trigger conditions via the application of only the pulse height threshold because the dual-gate trigger effectively \Erase{collected the events with energy deposit in the CaF$_2$ crystal.}\Add{removed the LS-events.}

The $\alpha$-events by $^{215}$Po decay ($^{235}$U series) contained within the CaF$_2$ crystal were used to create the reference pulse shape for $\alpha$-events because $^{215}$Po decay can be accurately selected via delayed coincidence analysis of $^{219}$Rn$\rightarrow$ $^{215}$Po (T$_{1/2}$ = 1.781~msec) $\rightarrow$ $^{211}$Pb. 
In addition, the $^{215}$Po decay was accompanied by almost no $\gamma$-rays.
Thus, a pure $\alpha$-event reference pulse shape was able to be obtained.

\subsubsection{PSD parameters}
Two types of PSD analysis were performed using each signal pulse shape, within 500~nsec from the starting time of the pulse shape, wherein the reference pulse shapes were particularly different among the $\alpha$-, $\beta$-, and $\beta$+LS-events.
The first analysis aimed to remove the $\beta$+LS-events, whereas the second analysis used a shape indicator~(SI)~\cite{ShapeIndicatorPRC67.014310} to discriminate between $\alpha$- and $\beta$-events in the CaF$_2$. 

We removed the $\beta$+LS-event by performing a chi-square test of each event pulse shape.
The pulse shapes observed with 62 PMTs were summed up first.
The 500~nsec to 4000~nsec interval from the starting time of the summed pulse shape was fitted using the reference pulse shape for $\beta$-events (i.e., normalizing the reference pulse shape to the pulse height of each event). 
The chi-square of each event pulse shape to the reference pulse shape was calculated in the 2~nsec time-bin region up to 500~nsec.
\Add{Errors in the pulse height were calculated based on statistical fluctuations in the number of photoelectrons in each time-bin, as shown in Fig.\ref{Fig:PulseShape}.}
The chi-square calculated using only the reference pulse shape for $\beta$-events was referred to as PSD$_{\beta}$, whereas that calculated using both the reference pulse shapes for $\beta$- and LS-events was denoted by PSD$_{\beta+\rm{LS}}$. 
At this point, the pulse height of the reference pulse shape for LS-events was also fitted such that PSD$_{\beta+\rm{LS}}$ was minimized, to evaluate the energy deposit in the LS.

\begin{figure}[ht]
\begin{center}
\includegraphics[width=1.0\linewidth]{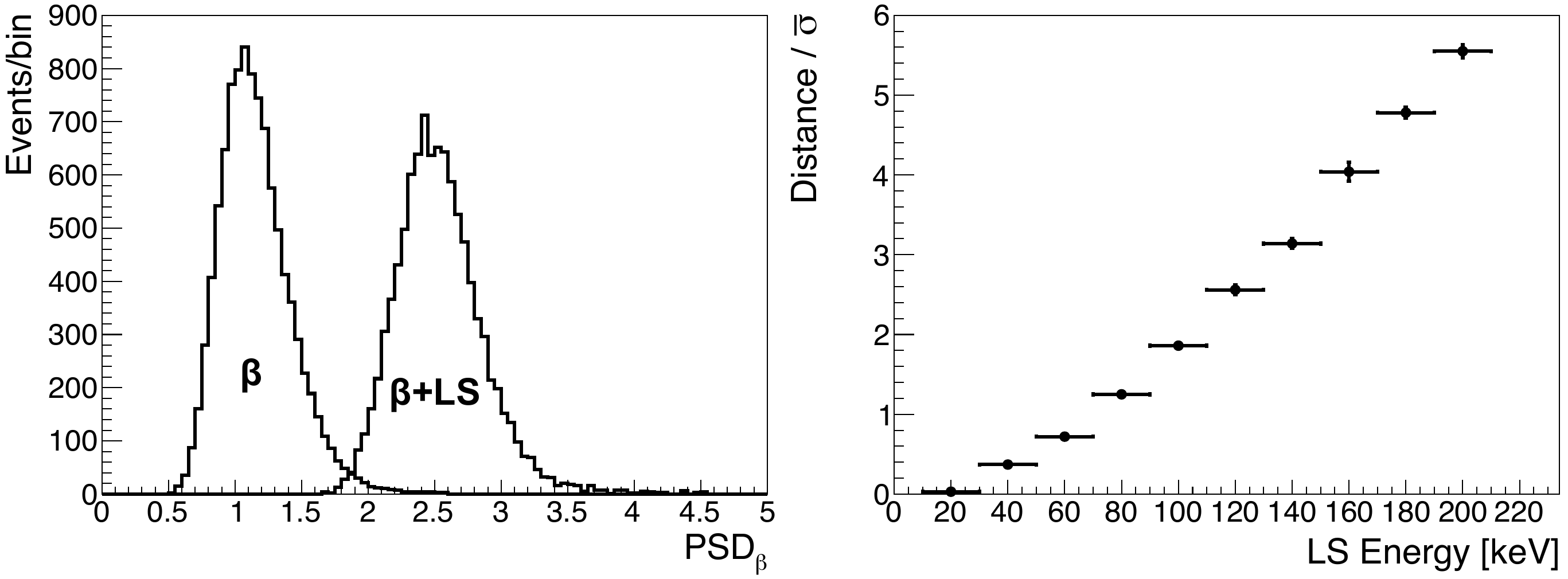}
\caption{
Separation of $\beta$- and $\beta$+LS-events via PSD analysis: 
(a) PSD$_\beta$ distribution of $\beta$- and $\beta$+LS-events. 
\Add{Pulse shapes of a 2.6\,MeV $\beta$-event and 140\,keV LS-event were randomly added to create pulse shape of $\beta$+LS-event.
2.6MeV $\beta$-event was the same as that used to create reference pulse shape of $\beta$-events.}
(b) LS energy dependence on discrimination ability obtained via the same analysis as in (a), at different LS energies.
}
\label{fig:PSD-beta_distribution}
\end{center}
\end{figure}

Fig.~\ref{fig:PSD-beta_distribution}(a) shows the PSD$_{\beta}$ distribution obtained for the $\beta$-events with energy deposits of 2.615~MeV and for the $\beta$+LS-events.
The pulse shape of the $\beta$+LS-events was artificially created via random combination of a 2.6\,MeV $\beta$-event and a 140\,keV LS-event.
The $\beta$-events peaked at 1 in the PSD$_\beta$ distribution, whereas the $\beta$+LS-events were distributed at large values, and vice versa, obtaining PSD$_{\beta+{\rm LS}}$.
Fig.~\ref{fig:PSD-beta_distribution}(b) shows the ability to discriminate between the $\beta$- and $\beta$+LS-events for each energy deposit in the LS at approximately 2.6\,MeV, where $\gamma$-rays from $^{208}$Tl decay were observed.
When the energy deposit of the LS was 140~keV (approximately 5\% of the total observed energy), the $\beta$+LS-events can be discriminated with a separation level of more than 3 $\overline{\sigma}$,
where $\overline{\sigma}$ is defined as $\sqrt{\sigma^{2}_{\beta}+\sigma^2_{\beta+{\rm LS}}}$.

For the PSD analysis, we adopted a shape indicator~(SI) to discriminate between $\alpha$- and $\beta$-events in the CaF$_2$.
\Add{The SI analysis reported the results of particle identification using ZnWO$_{4}$, CdWO$_{4}$ and CaWO$_{4}$ scintillators in Refs.\cite{ShapeIndicatorPRC67.014310},\cite{CdWO4PSD}, and \cite{CaWO4PSD}, respectively.
The SI was defined as
\begin{eqnarray}
\label{eq:si}
\rm{SI} = \it \frac{\Sigma (f(t_{k})- \overline{f}_{\beta}(t _{k})) \times P(t_{k}))}{\Sigma f(t_{k})},
\end{eqnarray}
where $f(t_{k})$ was the pulse height obtained at $t_{k}$,
and $\overline{f}_{\alpha}(t _{k})$ and $\overline{f}_{\beta}(t _{k})$ were reference pulse shapes for $\alpha$- and $\beta$-events, respectively.
On the other hand, the weight function $P(t)$ was defined as
\begin{eqnarray}
\label{eq:si-weight}
P(t) =  \frac{\overline{f}_{\alpha}(t)-\overline{f}_{\beta}(t)}
{\overline{f}_{\alpha}(t)+ \overline{f}_{\beta}(t)}.
\end{eqnarray}
}
The SI had a better discrimination ability when weighting was performed to emphasize pulse shape differences.
The $\alpha$-events were able to be discriminated from the $\beta$-events at the 4 $\sigma$ level in the 2.6\,MeV region,
which corresponds with an energy of 7.7~MeV for the $\alpha$-rays (Fig.~3 of Ref.~\cite{Umehara:2015lla}).

\subsection{\Add{Hit-crystal determination and} position reconstruction}
\label{sec:position}
The crystal determination of an event, wherein the event deposits energy (hit-crystal), is important for energy reconstruction and background rejection.
\Add{The hit-crystal was determined based on the position reconstructed with signals from 62 PMTs.}
The position was calculated via the light yield centroid method using the following formula:
\begin{equation}
\overrightarrow{\rm Position}  = \frac{ \sum^{62}_{i=1} \left(N_{pe}(i) \times \overrightarrow{\rm PMT}(i) \right)}{\sum^{62}_{i=1} N_{pe}(i)}.
\end{equation}
Here, $N_{pe}(i)$ is the number of photoelectrons observed in each PMT, and $\overrightarrow {\rm PMT}(i)$ denotes the position coordinates of each PMT.
\Add{The number of photoelectrons was calculated using the charge of a single photoelectron, which was evaluated by a random trigger event using a 3 Hz pulser.}
The pedestal was calculated in the first 180~nsec before the starting time of the pulse shape (Fig.~\ref{Fig:PulseShape}).
The pedestal was subtracted, integration was performed over the 4\,$\mu$sec range from the starting time to obtain the charge, and the value was converted to the number of photoelectrons detected by each PMT.
The photoelectrons were always detected even in the distant PMT for the events above the energy threshold (approximately 1\,MeV) since the number of detected photoelectrons by CANDLES-III was approximately 1000\,photoelectrons/MeV.

\begin{figure}[ht]
\begin{center}
\includegraphics[width=0.5\linewidth]{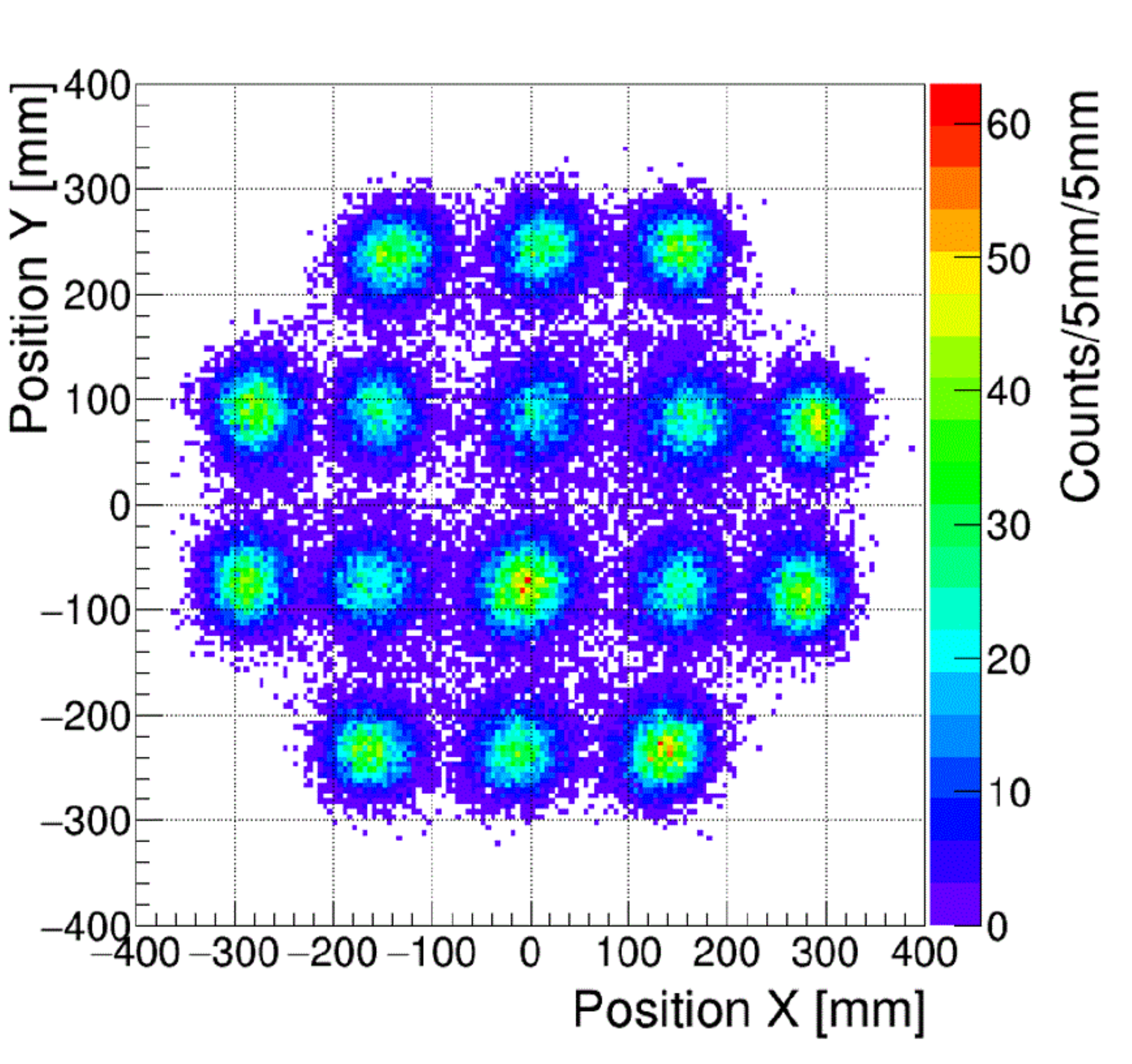}
\caption{Two-dimensional distribution of reconstructed position (X versus Y directions).
Actual distance to adjacent crystal is 20 cm, but reconstructed position is not corrected for that distance.} 
\label{Fig:Position}
\end{center}
\end{figure}

Fig.~\ref{Fig:Position} shows a two-dimensional plot of the reconstructed position \Add{of the 16 top-layer crystals} for the $\beta$-events caused by $\gamma$-rays from $^{40}$K decays contained mostly in the PMTs.
Each cluster corresponded to a crystal. 
The hit-crystal was then unambiguously determined.
Subsequently, Gaussian fitting was performed on three axes for each crystal.
Its mean and $\sigma$ values were then evaluated.
\Add{Based on the obtained mean and $\sigma$ values, the distance between the reconstructed crystal positions was approximately 6 $\sigma$.
This corresponds to 7 -- 8 $\sigma$ at the Q$_{\beta\beta}$-value region when the energy dependence was considered.
Thus, in CANDLES-III, hit-crystals were able to be selected with good accuracy.}
\Erase{Consequently, the peak-to-peak distance between the crystals was found to be 7--8 sigma.
The mean and sigma values obtained here were used in the analysis for the hit-crystal determination.}

\subsection{Energy reconstruction and energy calibration}
\label{sec:calib}
The amount of light yield for each CaF$_2$ module was slightly different from those of the others, 
and the light collection efficiency differed depending on the position.
The hit-crystal was determined based on the reconstructed position.
The energy for each event was then determined based on the number of photoelectrons, using the following formula:
\begin{equation}
{\rm Energy} = \sum^{62}_{i=1} N_{pe}(i) \times C_{relative}(N_m) \times C_{fine}(N_l).
\end{equation}
Here, $C_{relative}$ is a photoelectron-energy conversion coefficient determined for each CaF$_2$ module, and $C_{fine}$ is a fine correction coefficient for energy scale linearity in the high energy region determined for each layer.
$N_m$ and $N_l$ represent the module and layer numbers, respectively.

We performed a relative energy calibration for each CaF$_2$ module with a 1.836\,MeV $\gamma$-ray of an $^{88}$Y source to determine $C_{relative}(N_{m})$.
Coefficient $C_{fine}$ was estimated for the whole data period with the background events of a $^{208}$Tl $\gamma$-ray (2.615\,MeV).
The energy scale and resolution near the Q$_{\beta\beta}$-value were evaluated using 3 to 9\,MeV $\gamma$-rays emitted by the neutron captures on Si, Fe, and Ni nuclei using a $^{252}$Cf neutron source.

\begin{figure}[htbp]
\begin{center}
\includegraphics[width=0.65\linewidth]{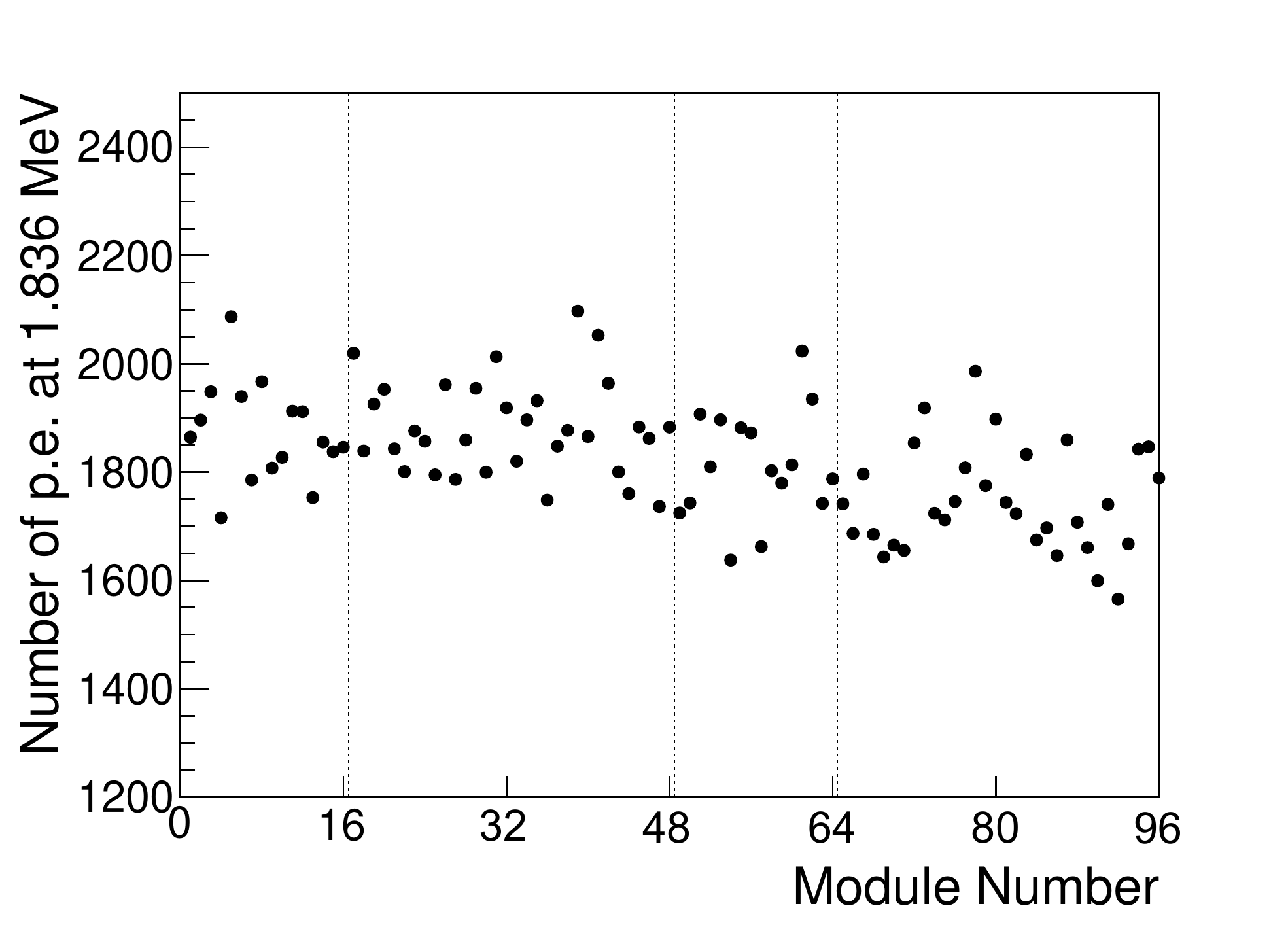}
\caption{Average number of photoelectrons for 1.836\,MeV $\gamma$-ray of $^{88}$Y as function of module number.
The 16 modules between two vertical dotted lines correspond to one layer. 
\label{Fig:Calib}}
\end{center}
\end{figure}

A calibration run was performed with the $^{88}$Y source inserted into the LS vessel and placed between the CaF$_2$ modules.
The data were collected at 18 locations in the detector to sufficiently irradiate all 96 modules with $\gamma$-rays.
Fig.~\ref{Fig:Calib} plots the average number of photoelectrons for each crystal when the 1.836\,MeV $\gamma$-ray was used. 
Accordingly, C$_{relative}$ was calculated to correct the variation.

The energy scale was then corrected using the background peak caused by 2.615\,MeV $\gamma$-rays from $^{208}$Tl decays.
Despite the relative correction between the CaF$_2$ modules,
a few percentage of the energy scale dependence existed on the Z direction of the detector.
The cause of the layer dependence was unknown, but it was considered herein to be the asymmetry in the Z direction of the detector (PMT sizes, etc.).
After this fine correction, the energy scale uniformity of all the crystals was confirmed \Add{to be within 0.3\%} using 1.461\,MeV $\gamma$-rays of $^{40}$K in the physics run data.

Finally, calibration was performed in the energy region above 3\,MeV using $\gamma$-rays emitted by the neutron capture reaction of $^{28}$Si and $^{58}$Ni.
This calibration was performed once per entire data taking term. 
Polyethylene bricks mixed with Si or Ni powder were created and assembled inside the Pb shield on top of the water tank.
A $^{252}$Cf source was placed in the center of the bricks to generate neutron capture $\gamma$-rays. 
The calibration results showed that the systematic error of the energy scale at the Q-value was less than 0.3\%.
\Erase{Furthermore, the}\Add{The} energy resolution at the Q$_{\beta\beta}$-value was estimated to be $\sigma$ = 2.4\%.
The details of the system and the analysis results were discussed in Ref.~\cite{Calib}.

After the calibrations were completed, the energy scale  stability was confirmed every 24 hours using the $^{208}$Tl peak.
The stability was determined to be better than 0.3\% for the live-time.

\section{Background analysis}
\label{sec:analysis}
In this chapter, we describe the detailed characteristics of the backgrounds relevant to CANDLES-III.
\Add{We accumulated experimental data for a live-time of 130.4 days to understand the origins of the backgrounds.
The strategies for reduction of background events are described in the following subsections.}

\subsection{Background candidates and reduction strategies} \label{sec:bg-candidate}
\subsubsection{(n,$\gamma$) reaction}\label{sec:n-gamma}
\Add{The environmental neutrons induce nuclear reactions leading to unstable nuclides, which then decay, emitting many $\gamma$-rays.}
In particular, prominent peaks were observed at 7 -- 8\,MeV in the initial run of CANDLES-III before the construction of the Pb shield.
\Add{The observed spectrum was well reproduced by simulated spectra of $\gamma$ rays from (n,$\gamma$) reactions, 
which originated from the surrounding rock (which abundantly contained Si and Fe isotopes) and a detector tank made of stainless steel (i.e., Fe, Ni, and Cr)~\cite{CANDLES-ngammaBG}.}
The environmental neutrons were induced by an ($\alpha$,n) reaction in the surrounding rocks.
The $\alpha$-rays were produced by decays of the progenies of $^{238}$U and $^{232}$Th content in the rocks~\cite{PTEP-neutron-Kamioka}.
A high energy $\gamma$-ray was occasionally absorbed by a single CaF$_{2}$ crystal because of the large size of the (10\,cm)$^{3}$ cube.
The neutron-induced background was reduced via the installation of the Pb shield outside of the water tank, as described in Section~\ref{sec:detector}.

The high energy $\gamma$-ray events were almost rejected by the PSD$_{\beta}$ analysis.
The remaining events caused by high energy $\gamma$-rays were estimated based on experimental data obtained 
by irradiation with an artificial neutron source.
The details regarding the (n,$\gamma$) background estimation were described in Ref.~\cite{CANDLES-ngammaBG}.
The background rate due to the neutron captured in the surrounding rocks was estimated to be 0.9$\pm$0.6 events/(96 CaF$_2$ crystals) with respect to the live-time.

\subsubsection{$^{208}$Tl decay} 
\label{sec:208Tl}
\begin{figure}[hbt]
\centering
\includegraphics[width=0.7\textwidth]{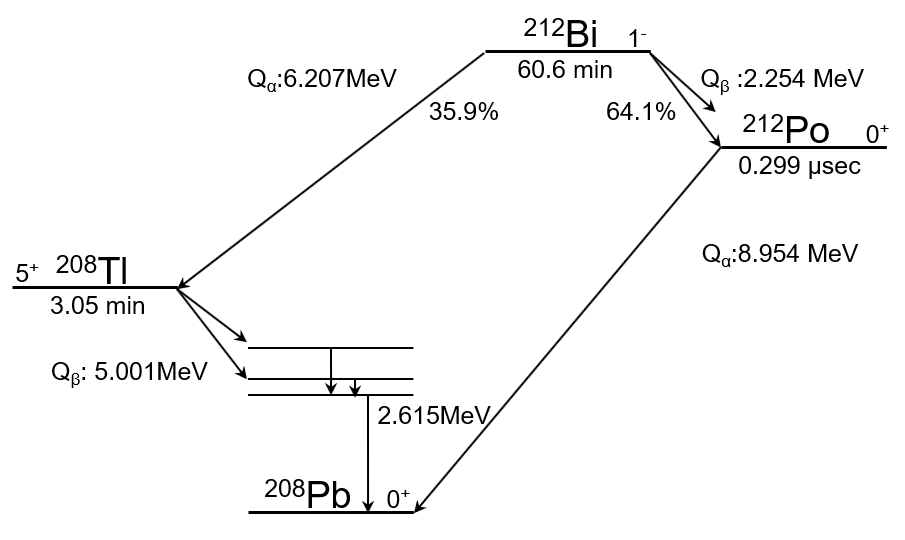}
\caption{Last part of $^{232}$Th series from $^{212}$Bi to $^{208}$Pb.
$^{208}$Tl always decays into the excited states of $^{208}$Pb, then decays into the ground state with 2.615\,MeV $\gamma$-ray emission.
}
\label{fg:212Bi}
\end{figure}

The $^{208}$Tl decay in the CaF$_{2}$ crystal was another background candidate for the 0$\nu\beta\beta$ decay search,
because the Q-value (5.001\,MeV) was above the Q$_{\beta\beta}$-value of $^{48}$Ca.
$^{208}$Tl did not directly decay to the ground state, but to the excited states of $^{208}$Pb, thereby always emitting 2.615\,MeV $\gamma$-rays (Fig.~\ref{fg:212Bi}).
Most of the events induced by $^{208}$Tl decay were removed by PSD$_\beta$ analysis
because the 2.615\,MeV $\gamma$-rays caused multiple scattering in both the CaF$_{2}$ and LS.
However, these events became a background when the $\beta$-ray and $\gamma$-ray (2.615\,MeV) from $^{208}$Tl decay were fully absorbed in the same CaF$_2$ crystal.
Such events can be effectively identified via the tagging of the preceding $\alpha$ decay of $^{212}$Bi $\rightarrow$ $^{208}$Tl.
The $\alpha$-ray of $^{212}$Bi was followed by the $\beta$ decay of $^{208}$Tl (T$_{1/2} = 3.05$~min).
The $\alpha$-ray with 6.05\,MeV energy was observed at 1.63~\Erase{MeV}\Add{MeV$_{\rm ee}$} in the energy scale determined by the energy calibration (Section~\ref{sec:calib}) because of the CaF$_{2}$ scintillation quenching.
When the $\beta$-ray and $\gamma$-ray (2.615~MeV) from $^{208}$Tl decay were absorbed by different CaF$_2$ crystals,
the event position was reconstructed in the LS area and not in the CaF$_2$ crystal.
These \Erase{muti-crystal}\Add{multi-crystal} events were rejected by position reconstruction analysis.

\subsubsection{$^{212}$BiPo event}
\label{sec:212BiPo}
The $\beta$ decays of $^{212}$Bi to $^{212}$Po \Add{(Q-value = 2.254\,MeV)} had a branching ratio of 64\% (Fig.~\ref{fg:212Bi}).
The half-life of $^{212}$Po (T$_{1/2}$ = 0.299\,$\mu$sec) was shorter than the decay-time of CaF$_{2}$ scintillation (1\,$\mu$sec); hence, the delayed $^{212}$Po $\alpha$ decay piled up the prompt $^{212}$Bi $\beta$ decay \Add{($^{212}$BiPo event)}.
\Erase{This sequential decay}\Add{The $^{212}$BiPo event} was observed as one event when the time-lag of the decays was relatively short.
The $^{212}$BiPo events were distributed in the energy region up to 5.1\,MeV~\cite{Umehara:2015lla}, since the $^{212}$Po $\alpha$-ray was observed at 2.88\,MeV$_{\rm ee}$ because of the quenching effect.
\Erase{The sequential decay of $^{212}$Bi$\rightarrow^{212}$Po $\rightarrow$ $^{208}$Pb ($^{212}$BiPo event) made the background for the $0\nu\beta\beta$ decay search.}

\subsection{Background rejection and 0$\nu\beta\beta$ decay analysis}
The criteria for selecting candidate events for the 0$\nu\beta\beta$ decay are given as follows:\\
(1) CaF$_2$ signal without energy deposit in LS;\\
(2) not a sequential signal caused by $^{212}$BiPo event;\\
(3) not a candidate for $^{208}$Tl decay; and\\
(4) reconstructed event at the CaF$_2$ crystal position.

As mentioned in section~\ref{sec:psd}, criterion (1) was applied using PSD analysis to remove $\beta$+LS-events.
\Add{The rejection efficiencies for $\beta$+LS-events at 2.6\,MeV were 77\% and 90\% when the energy deposits in LS were 50 and 60\,keV, respectively.}

\begin{figure}[tbh]
\begin{center}
\includegraphics[width=0.85\linewidth]{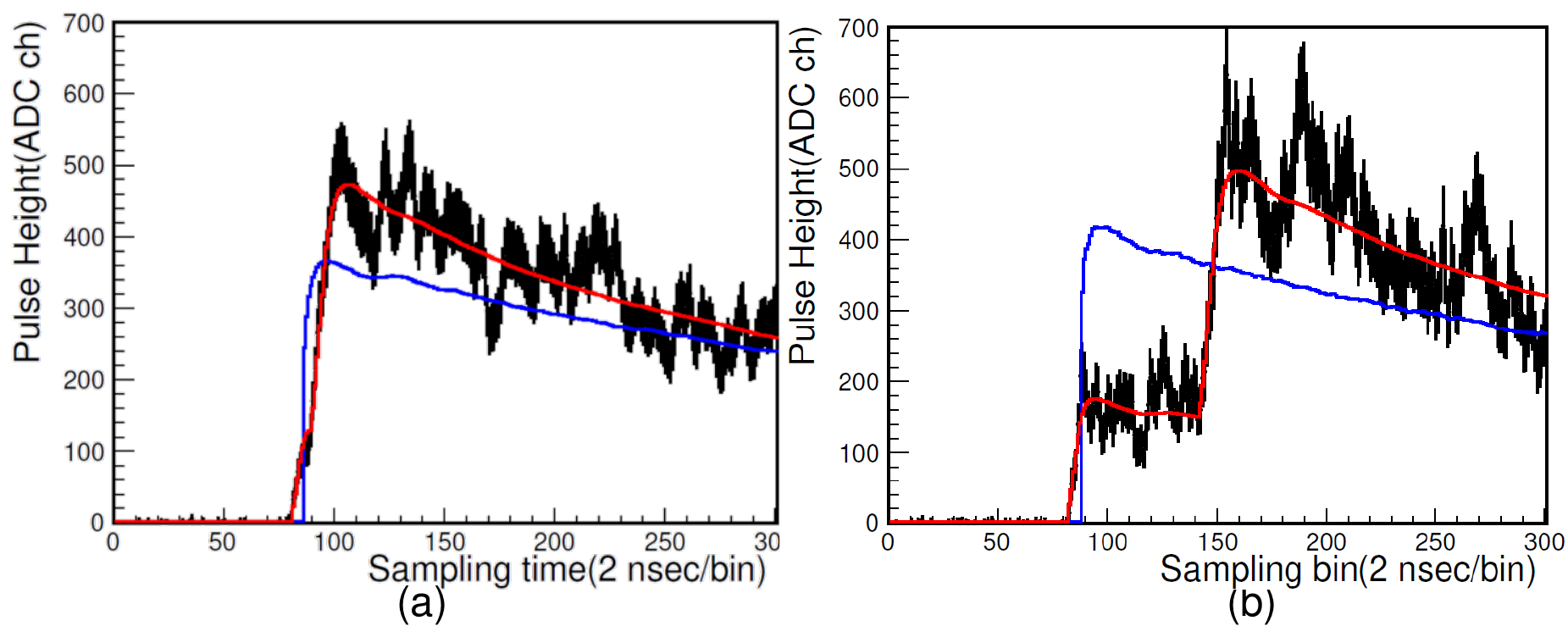}
\caption{
\Add{Typical pulse shapes of $^{212}$BiPo events. 
(a)Event with time-lag shorter than 10\,nsec.
Red and blue lines show reference pulse shapes for $\alpha$- and $\beta$-events, pulse heights of which are fitted by integrating pulse shapes between 500\,nsec and 4000\,nsec.
(b)Event at which time-lag is estimated to be 46\,nsec.
Red line shows fitting result with reference pulse shapes for $\beta$-event~(prompt) and $\alpha$-event~(delayed). 
Blue line shows reference pulse shapes for $\alpha$-event normalized as in (a).}
}
\label{Fig:212BiPo}
\end{center}
\end{figure}

Criterion (2) was applied via analysis of the pulse shape~\cite{Umehara:2015lla}.
\Add{
A typical $^{212}$BiPo event rejected by criterion (2) is shown in Fig.~\ref{Fig:212BiPo}(b).}
The $^{212}$BiPo event was easily recognized as a sequential event when the time-lag was longer than 10\,nsec since the typical rise-time of the CaF$_{2}$ signal was \Add{faster than} 10\,nsec.
Events with time-lags shorter than 10\,nsec were rejected using SI to discriminate between $\beta$- and $\alpha$-events, \Add{as shown in Fig.~\ref{Fig:212BiPo}(a),}
because a large portion of the pulse shape was caused by $^{212}$Po $\alpha$-events. 
The total rejection efficiency for $^{212}$BiPo events was more than 99\%.

Criterion (3) was applied via a time correlation analysis between $^{212}$Bi $\alpha$ decay and $^{208}$Tl $\beta$ decay.
The $^{212}$Bi $\alpha$-ray was identified using SI and its energy (1.63\,\Erase{MeV}\Add{MeV$_{\rm ee}$}).
The timing and the hit-crystal of the $\alpha$-events were recorded.
The event was observed in the identical crystal and within 18\,minutes after the $\alpha$-event was tagged as a $^{212}$Bi$\rightarrow^{208}$Tl$\rightarrow^{208}$Pb event~~\cite{Umehara2016}.
The rejection efficiency for $^{208}$Tl decays by this $\alpha$-tagging analysis was 89\%.

The event in which a 2.615\,MeV $\gamma$-ray was absorbed in the other CaF$_{2}$ crystal as a $^{208}$Tl $\beta$-ray can be rejected through criterion (4).
This multi-crystal event may be reconstructed in the position of the LS area.
To apply criterion (4), we selected the events within the $\pm 2~\sigma$ region from the center of each CaF$_{2}$ crystal, as presented in Section~\ref{sec:position}.

\begin{figure}[tbh]
\begin{center}
\includegraphics[width=1.0\linewidth]{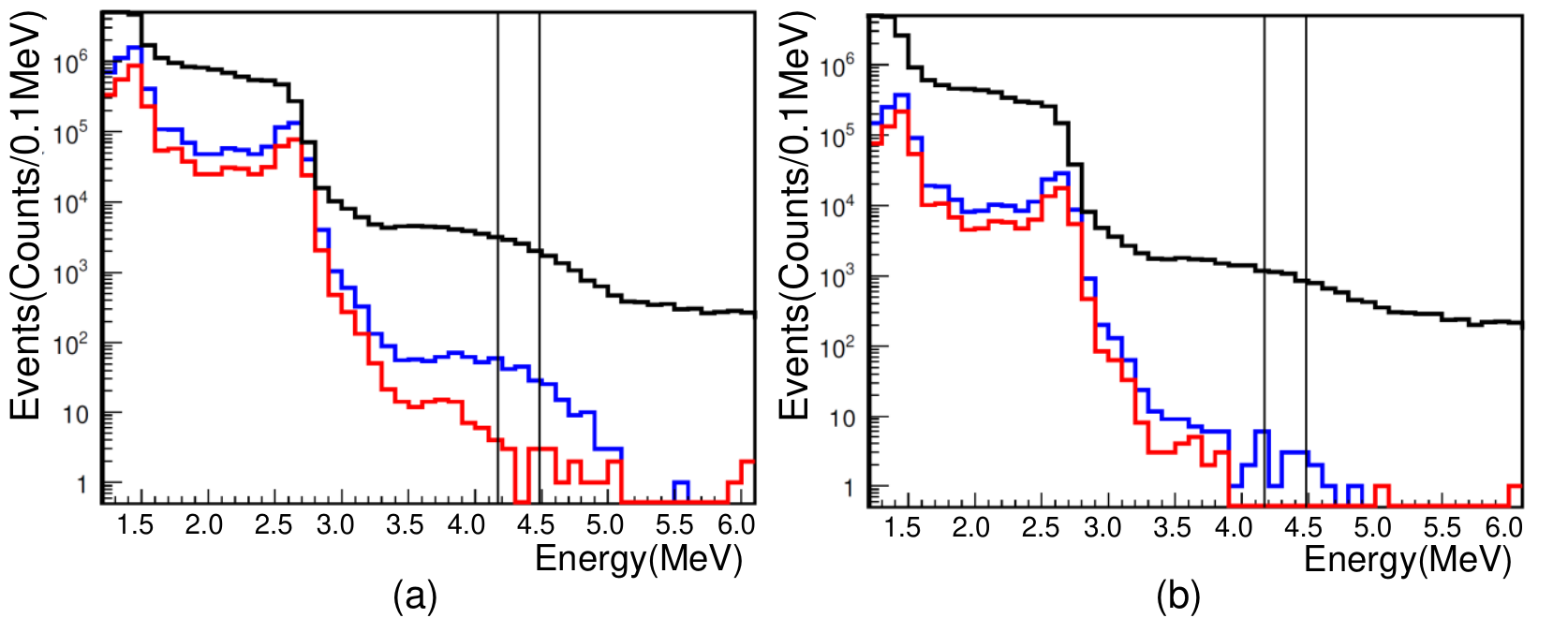}
\caption{Obtained energy spectra from each event selection using (a) 93 CaF$_{2}$ crystals and (b) 21 high purity CaF$_{2}$ crystals.
Black lines correspond to raw events without any selections.
Blue lines correspond to events after applying selection criteria (1) and (2).
Red lines correspond to events after applying criteria (1) -- (4).
Details of the event selection criteria (1) -- (4) are described in text.
The region between two vertical lines represents 0$\nu\beta\beta$ window of 4.17 -- 4.48\,MeV.
After event selections, no events are observed in the Q$_{\beta\beta}$-value region when 21 high purity crystals are selected.}
\label{Fig:Spectrum-27and93}
\end{center}
\end{figure}

\begin{figure}[ht]
\begin{center}
\includegraphics[width=0.98\linewidth]{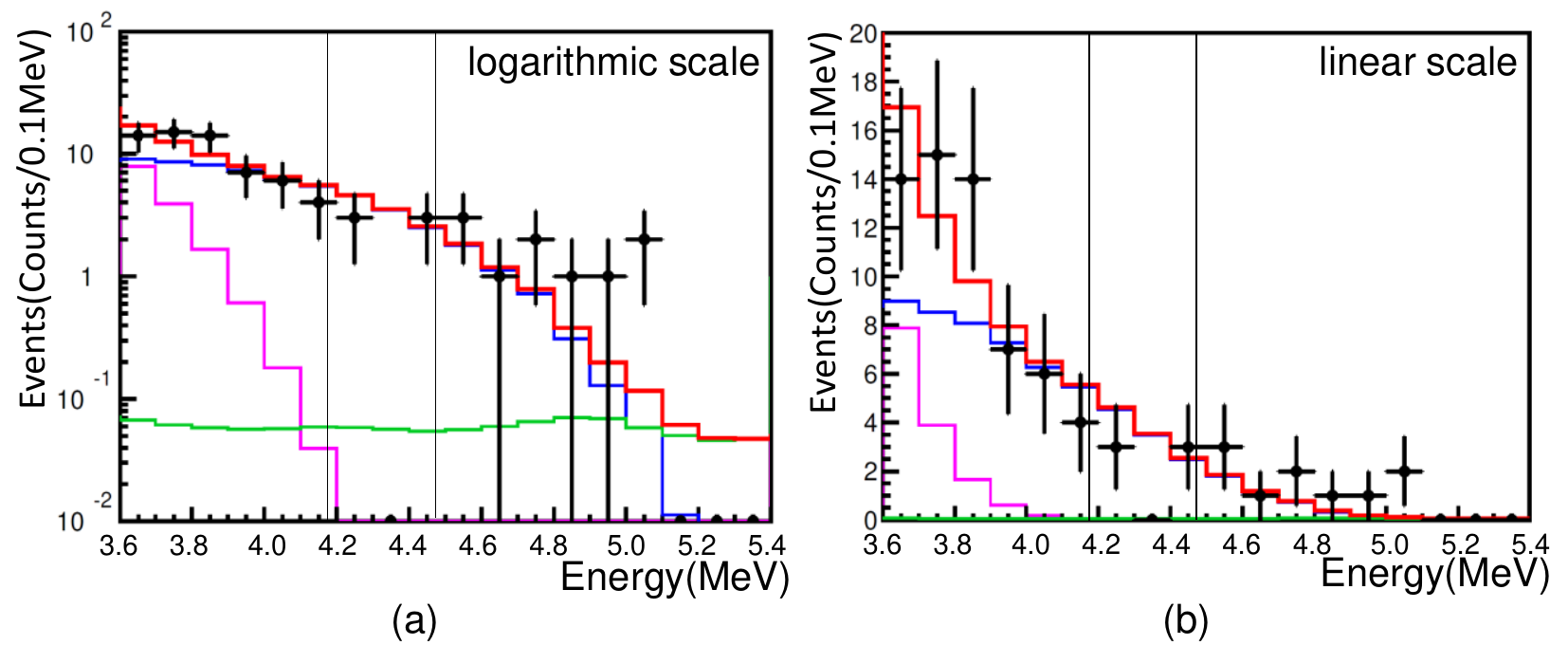}
\caption{Obtained energy spectra (plots) and simulated background spectra (lines) obtained using 93 CaF$_{2}$ crystals \Add{in (a) logarithmic scale and (b) linear scale}.
Blue, magenta, and green lines correspond to simulated background spectra of $^{208}$Tl and $^{212}$BiPo, 2$\nu\beta\beta$ decay, and (n,$\gamma$) events, respectively.
Red line represents summed up spectrum of all three simulated spectra.
The experimental energy spectrum in the Q$_{\beta\beta}$-value region is well reproduced by the simulated one. 
}
\label{Fig:SimSpectrum}
\end{center}
\end{figure}

Figs.~\ref{Fig:Spectrum-27and93}(a) and (b) show the energy spectra obtained by applying the selection criteria using 93 and 21 CaF$_{2}$ crystals, respectively.
We used 93 CaF$_{2}$ crystals for 0$\nu\beta\beta$ decay analysis because three out of the 96 CaF$_{2}$ crystals had poor performance.
\Add{The first excluded crystal, which was approximately 100 times more contaminated than the other crystals, was intentionally installed to evaluate the performance of PSD analysis and the stability of the detector. 
The second crystal had a quenching factor for $\alpha$ rays that was different from those of the other crystals,
and was therefore not used for the analysis because the pulse shapes of the $\alpha$-events may be different. 
The third crystal, on the other hand, exhibited a significant decrease in light yield during the measurement period.
For the aforementioned reasons, three crystals out of 96 were not used in the analysis.
}
Meanwhile, 21 CaF$_{2}$ crystals were selected as high purity crystals, with radioactive impurities of the $^{232}$Th series that were less than 10\,$\mu$Bq/kg.
\Add{The reasons for setting this impurity level for high-purity crystals will be discussed in detail in the next section.}
The blue lines in the figures correspond to the events relevant to criteria (1) and (2), which were required for rejecting $\beta$+LS-events.
The event rates of the blue lines were less than those of the black lines by more than two orders of magnitude above 3.5\,MeV, indicating that the LS efficiently worked as a 4$\pi$ active shield for the CaF$_2$ crystals to reduce the external backgrounds.
The red spectra in Fig.~\ref{Fig:Spectrum-27and93} were obtained by applying event selection criteria (1) -- (4).

\begin{table*}
\caption{\label{tb:0nbb_effi}
\New{Summary of detection efficiencies.}}
\begin{ruledtabular}
\begin{tabular}{l|cc}
\makebox[100mm]{   } &
\makebox[30mm]{93 crystals} &
\makebox[30mm]{21 crystals}\\\hline  
(a) criteria (1)+(2)                 &  \multicolumn{2}{c}{0.715}        \\
(b) criteria (1)+(2)+(4)             &  \multicolumn{2}{c}{0.709}       \\
(c) energy range(4.17--4.48MeV) + criteria (1)+(2)+(4) &  \multicolumn{2}{c}{0.493}  \\
(d) criterion (3)($^{208}$Tl cut)    & 0.727 & 0.761      \\
(e) total efficiency= (c)$\times$(d) & 0.358 & 0.375      \\
\end{tabular}
\end{ruledtabular}
\end{table*}

\New{
The detection efficiency\Erase{which includes the acceptance efficiency} for 0$\nu\beta\beta$ decay after applying \Add{selection} criteria (1) -- (4) was evaluated by a Monte Carlo~(MC) simulation.
The detection efficiency after applying criteria (1)+(2)+(4) was estimated to be 70.9\%.\Erase{ for the 93 CaF$_2$ crystals.}
The efficiency was reduced to 49.3\% by selecting events in the 0$\nu\beta\beta$ window, as listed in Tab.\ref{tb:0nbb_effi}(c).
The efficiency~(c) was reduced by criterion~(3), which vetoed the events within 18 minutes after the $^{212}$Bi $\alpha$-rays candidate events.
\Add{As listed in Tab.\ref{tb:0nbb_effi}(d), the live-times of the data were reduced to 72.7\% and 76.1\% for the 93 and 21 crystals, respectively, because the total vetoed time was dependent on the radioactive impurities in the CaF$_{2}$ crystals.}
Finally we obtained the total detection efficiency for the 0$\nu\beta\beta$ events, as listed in Tab.\ref{tb:0nbb_effi}(e).
}

\New{
\Erase{Here we discuss systematic errors.}
\Erase{These}\Add{The systematic errors} are mainly from uncertainties
in the \Erase{estimation of} following \Erase{three}\Add{4} items:\\
\Erase{
\begin{enumerate}
    \item {Uncertainty on absolute energy calibration and gain stability\cite{Calib} may 
    obscure the 0$\nu\beta\beta$ window. We found it to be less than 2.5 \%.}
    \item {Uncertainty on PSD efficiencies, which include uncertainty from LS calibration was estimated to be less than 0.5 \%.}
    \item {Uncertainty on radioactivities in the CaF$_{2}$ crystals
    may change the estimation of backgrounds.   
    It was estimated to be less than 2.5 \%.}
    \item {Uncertainty on LS calibration
    may change the estimation of backgrounds.
    It was estimated to be 15 \%.}
\end{enumerate}
}
(1)Uncertainty in absolute energy calibration and gain stability~\cite{Calib}, which may obscure the 0$\nu\beta\beta$ window, was determined to be less than 2.5\%.\\
(2)Uncertainty in PSD efficiencies\Erase{, which include uncertainty from LS calibration} was estimated to be less than 0.5 \%.\\
(3)Uncertainty in radioactivities in the CaF$_{2}$ crystals, which may change the estimation of backgrounds, was estimated to be less than 2.5\%.\\
(4)Uncertainty in \Erase{LS calibration}\Add{energy dependence of the LS light yield}, which may change the estimation of backgrounds, was \Add{conservatively} estimated to be 15\%.\\
}
\New{
Uncertainties (1), (2), and (3) were much smaller than the statistical error,
whereas uncertainty (4) was relatively large\Erase{ comparing with statistical error}. 
\Erase{Thus}\Add{Therefore,} we \Erase{take}\Add{considered} uncertainty (4) in deriving the half-life limit.}

\Erase{The detection efficiencies of the 0$\nu\beta\beta$ decay applying PSD analyses were estimated to be 70.9\% and 71.8\% for the 93 and 21 CaF$_2$ crystals, respectively.
These efficiencies were reduced to 49.3\% by selecting events in the 0$\nu\beta\beta$ window for both CaF$_2$ crystal cases.
The events within 18 minutes after the $^{212}$Bi $\alpha$-rays candidate detection were rejected by applying criterion (3).
The live-time of the data was reduced to 72.7\% and 76.1\% for the 93 and 21 CaF$_2$ crystals, respectively.}

Fig.~\ref{Fig:SimSpectrum} shows the simulated background and measured spectra for the 93 CaF$_{2}$ crystals by applying event selection criteria (1) – (4).
The red-colored spectrum represents the sum of the simulated background spectra, comprising $^{212}$BiPo events, $^{208}$Tl decays, $\gamma$-rays by neutron capture reactions, and 2$\nu\beta\beta$ decays with a half-life of T$_{1/2}^{2\nu\beta\beta} = 5.3 \times 10^{19}$ year~\cite{Barabash2019}.
We considered the following parameters for the background rate estimation: 
(1) concentration of the radioactive impurities of the $^{232}$Th series in each CaF$_{2}$ crystal, determined via time-correlation analysis of the decay $^{220}$Rn $\rightarrow$ $^{216}$Po (T$_{1/2}$ = 145\,msec) $\rightarrow$ $^{212}$Pb;
(2) detection efficiency of the event selection criteria; and 
(3) detector energy resolution~\cite{Calib}.
We obtained an estimated background rate of 27.1 counts/130.4 days/(93 CaF$_{2}$ crystals) within the 4--5\,MeV region (Fig.~\ref{Fig:SimSpectrum}). 
The estimated background rate was consistent with the measured rate for 24 events.
This observation strongly supported our hypothesis that the three investigated background candidates were major candidates for the CANDLES-III detector.

\Erase{We can set a lower limit and an experimental sensitivity on the half-life of the 0$\nu\beta\beta$ decay by using the expected background rate of 1.0 counts in the 0$\nu\beta\beta$ window of 4.17 -- 4.48 MeV for the 21 CaF$_{2}$ crystals.
The half-life limit with 90\% C.L. obtained by selecting 21 CaF$_{2}$ crystals was 5.6 $\times$ 10$^{22}$ year.
This limit was comparable to the result obtained for more than 2 years using our previous detector, ELEGANT VI\cite{Umehara2008}.
We also obtained an experimental sensitivity of 2.7 $\times$ 10$^{22}$ year (90\% C.L.) because the observed event rate was lower than expected.
The obtained half-life limit derived an upper limit on the effective Majorana neutrino mass $\langle m_{\nu}\rangle\le$ 2.9 -- 16~eV (90\% C.L.) using the nuclear matrix elements given in Ref.~\cite{Engel_2017} and the reference therein. }

\Erase{The present limits on the half-life and the effective Majorana mass were obtained using natural Ca instead of enriched $^{48}$Ca crystals.
The limit on $\langle m_{\nu} \rangle$ did not reach sufficient sensitivity compared with the experiments using other enriched $\beta\beta$ isotopes, such as $^{76}$Ge and $^{136}$Xe, because of the lack of $^{48}$Ca isotope amount overcome by realizing $^{48}$Ca enrichment. 
The results obtained herein demonstrated that $^{48}$Ca is a promising isotope that is sufficiently competitive in other sensitive experiments.}

\section{Discussion and Perspectives}
\Add{The  major  backgrounds  in the Q-value region of CANDLES-III were the $^{208}$Tl and $^{212}$BiPo decays within the CaF$_2$ crystals.
Therefore, the amount of $^{232}$Th series impurities was used as an index for the high-purity crystal.
According to Fig.~\ref{Fig:Spectrum-27and93}(b), 21 CaF$_{2}$ crystals were selected as high purity crystals, with radioactive impurities of the $^{232}$Th series that were less than 10~$\mu$Bq/kg.
Based on the MC simulation, the expected background in the Q-value region was estimated to be 1 event for a live-time of 130.4 days, for the 21 CaF$_2$ crystals..
In addition, we have been developing high-purity CaF$_2$ crystals.
The amounts of $^{232}$Th series impurities in recently produced 14 CaF$_2$ crystals were determined to be less than 10 $\mu$Bq/kg.
Thus, technology for producing high purity CaF$_2$ crystals containing impurities less than 10$\mu$Bq/kg has been established.
We can replace relatively contaminated CaF$_2$ crystals with these high purity crystals.
Consequently, we can extend low background measurement in the future.}

\Add{We observed no events in the 0$\nu\beta\beta$ window 4.17 -- 4.48\,MeV for the selected 21 high purity CaF$_{2}$ crystals, whereas \Erase{six}\Add{6} events were observed for the 93 CaF$_{2}$ crystals.
According to the MC simulation, the expected background rate was estimated to be 1.0 counts in the 0$\nu\beta\beta$ window for the 21 CaF$_{2}$ crystals, considering the position of the crystals and the impurities contained therein.
We can set a lower limit and experimental sensitivity on the half-life of 0$\nu\beta\beta$ decay.
The half-life limit with 90\% C.L. obtained by selecting 21 high purity CaF$_{2}$ crystals was 5.6 $\times$ 10$^{22}$ year.}
\Add{This limit was comparable to the most stringent value obtained via measurement for over two years using the ELEGANT VI detector~\cite{Umehara2008}.}
\Erase{This limit was comparable to the result obtained for more than 2 years using our previous detector, ELEGANT VI\cite{Umehara2008}.}
\Add{We also obtained an experimental sensitivity of 2.7 $\times$ 10$^{22}$ year (90\% C.L.)\Erase{ because the observed event rate was lower than expected}.
Based on the obtained half-life limit, the upper limit on the effective Majorana neutrino mass $\langle m_{\nu}\rangle\le$ 2.9 -- 16~eV (90\% C.L.) was derived using the nuclear matrix elements obtained from Ref.~\cite{Engel_2017} and the reference therein.}

\Add{The present limits on the half-life and effective Majorana mass were obtained using $^{\rm nat.}$Ca instead of enriched $^{48}$Ca crystals.
The limit on $\langle m_{\nu} \rangle$ did not reach sufficient sensitivity compared with those in experiments using other enriched $\beta\beta$ isotopes, such as $^{76}$Ge and $^{136}$Xe, because of the lack in $^{48}$Ca isotope.
The natural abundance of $^{48}$Ca is approximately 0.2\%, which is the smallest among the $\beta\beta$ decaying nuclei used in other experiments.
On the other hand, this result exhibits \Erase{a big advantage}\Add{a large potential} of approximately 500 enhancement in case enrichment could be achieved.
\Add{The highest enrichment achieved thus far is 96.6\%, which leads to approximately 540 enhancement~\cite{48Ca1966}.}
Several approaches to $^{48}$Ca enrichment, \Erase{including}\Add{such as} electromagnetic/optical separators~\cite{Elemag2001}, 
laser isotope separation~\cite{LIS2020}, 
chemical separation~\cite{Umehara-PTEP2015}~\cite{Hazama-KURRI}, 
and electrophoresis~\cite{Kishimoto-PTEP2015},
each with the aim of producing large amounts of $^{48}$Ca, are under development.}

\Add{CANDLES-III has achieved a 300\,kg large-volume detector and low background measurement.
The observed event rates in the 0$\nu\beta\beta$ window, considering the detection efficiency, was 1.0$\pm$0.4$\times$10$^{-3}$~events/keV/yr/(kg of $^{\rm nat.}$Ca) for the 93 CaF$_{2}$ crystals, which was comparable or less than those of other sensitive experiments.}
The results obtained herein demonstrated that $^{48}$Ca is a promising isotope that is sufficiently competitive \Erase{in}\Add{for} other sensitive experiments.

\section{Conclusion}
This study evaluated the feasibility of low background measurements with the CANDLES-III detector using data for a live-time of 130.4~days.
We confirmed that the structure of the 4$\pi$ active shield and passive shields can effectively reduce external backgrounds.
Backgrounds caused by radioactive impurities of the $^{232}$Th series contained in the CaF$_2$ crystals can be \Erase{effectively removed} \Add{reduced} through the analysis of the signal pulse shape and tagging of the time-correlated $\alpha$ decay.
After background rejection analyses, and when we selected 21 high purity CaF$_{2}$ crystals, no events in the Q$_{\beta\beta}$-value region were detected.
\Erase{This gave a lower limit on the half-life of the 0$\nu\beta\beta$ decay of $^{48}$Ca as T$^{0\nu\beta\beta}_{1/2}$ $\textgreater$ 5.6 $\times$ 10$^{22}$ yr (90\% C.L.), which was almost comparable with the limit obtained by ELEGANT VI over two years of long-term data~\cite{Umehara2008}.
We also presented the experimental sensitivity of 2.7 $\times$ 10$^{22}$ yr (90\% C.L.) by the predicted background rate.}
The observed energy spectrum around the Q$_{\beta\beta}$-value region was well reproduced by the simulated one, estimated with the three background candidates considered.
In other words, there were likely no additional high-impact backgrounds.
\Add{The observed event rate in the 0$\nu\beta\beta$ window was approximately 10$^{-3}$~events/keV/yr/(kg of $^{\rm nat.}$Ca), which was comparable or less than those of other sensitive 0$\nu\beta\beta$ experiments.}
The present result is useful for the development of a more advanced detector and shows that $^{48}$Ca is a promising target nucleus for the 0$\nu\beta\beta$ decay search using CaF$_2$ crystals.

\begin{acknowledgments}
This work was supported by JSPS/MEXT KAKENHI Grant Number 19H05804, 19H05809, 26104003, 16H00870, 24224007, and 26105513.
This work was supported by the research project of Research Center for Nuclear Physics~(RCNP), Osaka University. 
This work was also supported by the joint research program of the Institute of Cosmic Ray Research~(ICRR), the University of Tokyo.
The work of KT was supported by JSPS Research Fellowship for Young Scientists.
The Kamioka Mining and Smelting Company has provided service for activities in the mine.
\end{acknowledgments}

\bibliography{CANDLES_BB}

\end{document}